\documentclass[aps,prl,twocolumn,superscriptaddress,10pt]{revtex4-1}  
\usepackage{graphicx}  
\usepackage{dcolumn}   
\usepackage{bm}        
\usepackage{amssymb}   
\usepackage[utf8x]{inputenc}
\usepackage{amssymb,amsmath}
\usepackage{gensymb}
\usepackage{textcomp}
\usepackage{lineno}
\usepackage[normalem]{ulem}

\graphicspath{{figures/}}

\usepackage{xcolor}

\begin{document}

\widetext


\title{Segregation of large particles in dense granular flows: A granular Saffman effect?}


\author{K. van der Vaart}
\affiliation{Environmental Hydraulics Laboratory, \'{E}cole Polytechnique F\'{e}d\'{e}rale de Lausanne, \'{E}cublens, 1015 Lausanne, Switzerland}
\affiliation{Multi-Scale Mechanics, ET and MESA+, University of Twente, P.O. Box 217, 7500AE Enschede, The Netherlands}

\author{M. P. van Schrojenstein Lantman}
\affiliation{Multi-Scale Mechanics, ET and MESA+, University of Twente, P.O. Box 217, 7500AE Enschede, The Netherlands}

\author{T. Weinhart}
\affiliation{Multi-Scale Mechanics, ET and MESA+, University of Twente, P.O. Box 217, 7500AE Enschede, The Netherlands}

\author{S. Luding}
\affiliation{Multi-Scale Mechanics, ET and MESA+, University of Twente, P.O. Box 217, 7500AE Enschede, The Netherlands}

\author{C. Ancey}
\affiliation{Environmental Hydraulics Laboratory, \'{E}cole Polytechnique F\'{e}d\'{e}rale de Lausanne, \'{E}cublens, 1015 Lausanne, Switzerland}

\author{A.R. Thornton}
\affiliation{Multi-Scale Mechanics, ET and MESA+, University of Twente, P.O. Box 217, 7500AE Enschede, The Netherlands}

\date{\today}


\begin{abstract}


We report on the scaling between the lift force and the velocity lag experienced by a single particle of different size in a monodisperse dense granular chute flow. The similarity of this scaling to the Saffman lift force in (micro) fluids, suggests an inertial origin for the lift force responsible for segregation of (isolated, large) intruders in dense granular flows. We also observe an anisotropic pressure/stress field surrounding the particle, which potentially lies at the origin of the velocity lag. These findings are relevant for modelling and theoretical predictions of particle-size segregation. At the same time, the suggested interplay between polydispersity and inertial effects in dense granular flows with stress- and strain-gradients, implies striking new parallels between fluids, suspensions and granular flows with wide application perspectives.

\end{abstract}
\pacs{.....}

\maketitle

Size-polydispersity is intrinsic to non-equilibrium systems like granular materials~\cite{aranson2006}. It gives them the ability to size-segregate when agitated, a process which spatially separates different sized grains~\cite{williams1976, rosato1987, ottino2000}, but is different from phase separation in classical fluids. Particle-size segregation in dense granular flows~\cite{midi2004, jop2006constitutive} has been intensively studied~\cite[e.g.][]{golick_daniels2009, fan2010, fan2011, drahun1983, savage_lun1988, windows2016, hill2014, weinhart2013, staron2015, tunuguntla2016, wiederseiner2011a, staron2014, harrington2013, staron2016, edwards2016, jing2017}, but a fundamental question remains unanswered: \emph{why} do large particles segregate?

It is generally understood that in dense granular flows both small and large particles are pushed away from high shear regions~\cite{fan2010, fan2011} or pulled by gravity~\cite{drahun1983, savage_lun1988}. The reason for the separation of the two species is that small particles move more effectively; they can carry proportionally more of the kinetic energy~\cite{hill2014, weinhart2013, staron2015, tunuguntla2016} and are also more likely to move into the gaps between larger particles; a process referred to as kinetic sieving~\cite{drahun1983, savage_lun1988}. Unfortunately, the concept of kinetic sieving breaks down when the large-particle concentration is very low. Particle-size segregation also occurs in situations without shear, possibly extremely slowly, by small particles moving under larger ones, which allows for a purely geometrical interpretation \cite{duran1993, dippel1995}. Other alternative explanations for segregation, such as wall-induced segregation \cite{knight1993}, do not apply in chute flows as studied here.

Current models for size-segregation in dense granular flows perform well when the small and large-particle concentrations are nearly equal~\cite{gray_thornton2005, thornton2006, fan2011, marks2012, tunuguntla2014}. When accounting for the effect of size-segregation asymmetry~\cite{vandervaart2015, gajjar2014}, models have been extended to more unequal concentrations, but they remain inaccurate in the limit of low large-particle concentrations. Extending models to this limit is critical because during segregation, and even after reaching a steady state, regions of low large-particle concentration occur and can persist throughout the flow~\cite{vandervaart2015, staron2016, jing2017}. Moreover, current models are either completely or partly phenomenological. Thus, to advance modelling, we should aim to understand the physical origin of segregation to derive the free state-variables from their microscopic quantities. An important related issue is that current constitutive models for dense granular flows only work with an average particle size~\cite{kamrin2012, henann2013}. If we are to implement size-distributions in these models a better understanding of micro-scale effects between large and small particles seems crucial.

In contrast to particle-size segregation, particle migration in suspensions, in the limit of low concentrations, is generally well understood~\cite[e.g.][]{phillips1992, boyer2011}. Arguably this progress has been aided by the fact that the fluid forces acting on a particle can be calculated, which can not be said for granular media. This inspired us to treat the particles that surround an intruder as a continuum and attempt to understand the forces acting on a segregating particle based on the measured continuum fields.

\begin{figure}[b]
\resizebox{\columnwidth}{!}{\includegraphics{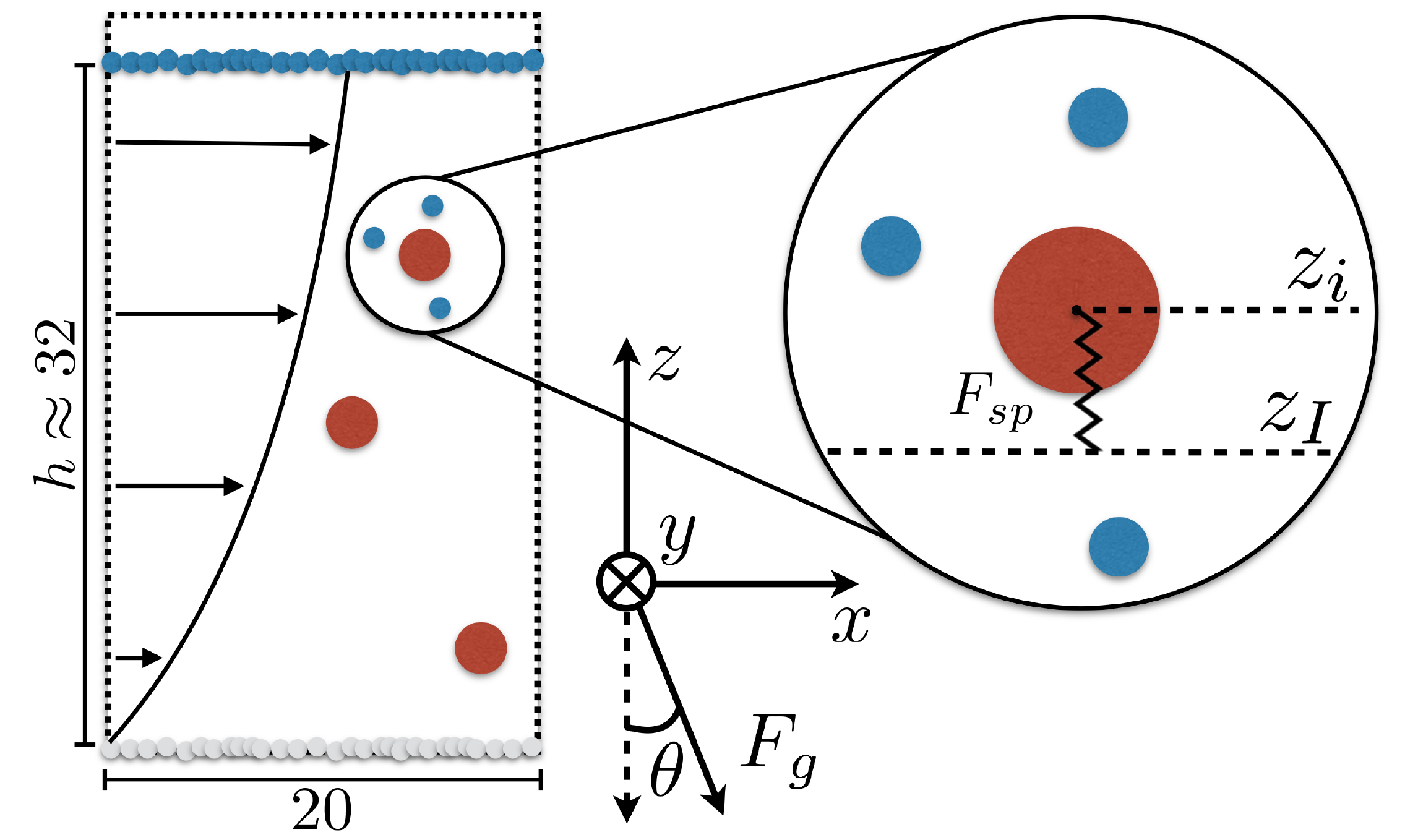}}
\caption{Schematic of the simulations: 3D mono-disperse granular flow down an incline, with angle $\theta = 22^{\circ}$. Only base (white) and surface (blue) particles are shown, as well as three bulk particles. The flow contains three intruder particles that are held with springs around three different $z$-positions $z_I$ (intruder positions in the schematic are to scale), but move freely in the $x$-$y$ plane.}
\label{fig:setup}
\end{figure}

 \begin{figure*}
\resizebox{\textwidth}{!}{\includegraphics{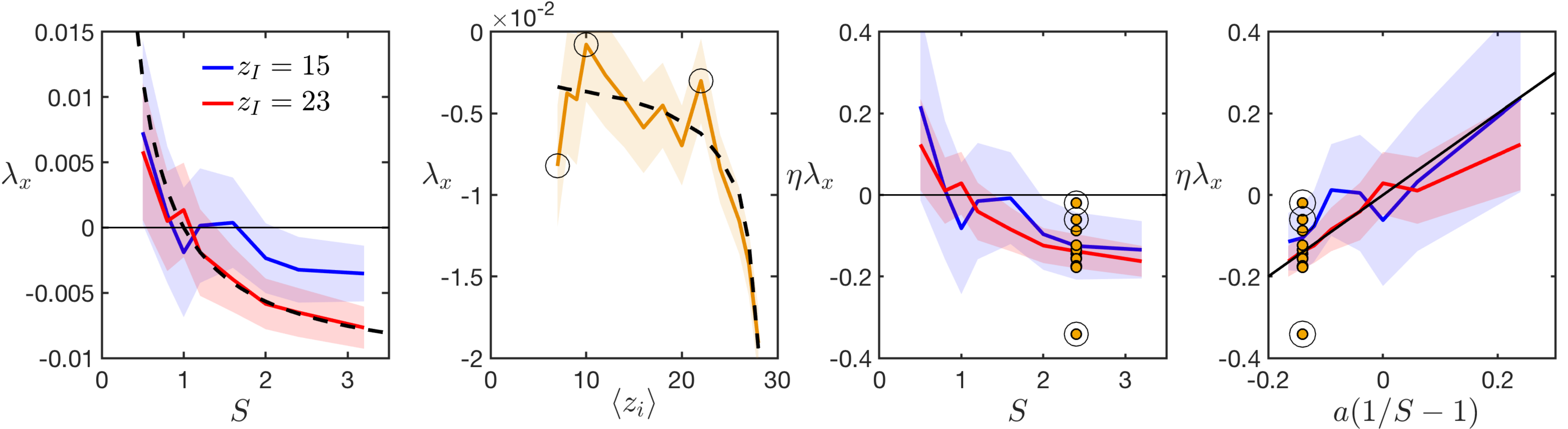}} 
\caption{($a$) The velocity lag $\lambda_x$ of the intruder particle as a function of size ratio $S$, for $z_I=15$ and $z_I=23$. ($b$) Velocity lag as a function of the average vertical position $\langle z_i \rangle$ of an intruder for $S=2.4$. The dashed lines in ($a$) and ($b$) are fits of $\lambda_x = a(\frac{1}{S} - 1)/\eta$, with $a=0.24$. The circles indicate the outliers. ($c$) The data from ($a$) and ($b$) are plotted here as $\eta\lambda_x$ versus $S$. The yellow circles are the data from ($b$), with the circles indicating the outliers. ($d$) The data from ($a$) and ($b$) are plotted here as $\eta\lambda_x$ versus $a(1/S -1)$. The solid black line has a slope of $1.0$. The yellow circles are again the data from ($b$), with the circles indicating the outliers.}
\label{fig:SvsLag_springParticle}
\end{figure*}

Recently, \citet{guillard2016} measured for the first time the segregation lift force on a single large intruder particle in a mono-disperse granular flow by attaching the intruder to a spring perpendicular to the plane (see Fig.~\ref{fig:setup}). They found scaling laws that linked the total upward force or net contact force on the intruder to shear and pressure gradients. These scaling laws predict the direction of segregation of large particles in different flow configurations depending on whether a shear or pressure gradient has the strongest contribution. However, they do not shed light on the origin of the lift force. 

In this study we present new physical insights into the origin of the segregation lift force on large intruders in three-dimensional mono-disperse dense granular flows. We do so, firstly, by taking a different approach to \citet{guillard2016} and determine the lift force $F_L$ by decomposing the net contact force on an intruder as $F_{c} = F_L + F_b$, where $F_b$ is a generalized buoyancy force for dense granular media that accounts for the local geometry around an intruder. This novel approach is inspired by our finding of an anisotropic pressure field that surrounds the intruder and grows with its size. Secondly, we report on a velocity lag of the intruder relative to the bulk flow and demonstrate a scaling between this velocity lag and the lift force. The similarity of this scaling to the known Saffman lift force in fluids and the presence of the anisotropic pressure field, allow us to propose the physical origin for the segregation lift force.

\emph{Methods.---}We use MercuryDPM, based on discrete particle methods (MercuryDPM.org;~\cite{thornton2013review,weinhart2017mercurydpm}), and investigate three-dimensional (3D) flows of mixtures of spherical dry frictional particles flowing down an incline of~$\theta = 22\degree$. We verified that changing the inclination angle between $22\degree$ and $26\degree$ has no significant effect (within the fluctuations) on the measured lift force $F_L$ (see the supplementary material). All simulation parameters are non-dimensionalized such that the particle density is $\rho_p = 6/ \pi$ and the gravitational acceleration is $g=1$, with vertical component $g_z = \cos{\theta}$. The simulations are conducted in a box with dimensions $(x,y,z)=(30, 8.9, 40)$, with periodic walls in the $x$ and $y$ directions. The particles that make up the bulk of the flow have a diameter $d_b = 1$.  We vary the intruder diameter $d_i$ between size ratios $S=d_i/d_b=0.5$ and $3.2$. The rough base consists of particles of diameter $0.85$ and the flow height is $h= 32 \pm 0.5$. 

\begin{figure*}
\resizebox{\textwidth}{!}{  \includegraphics{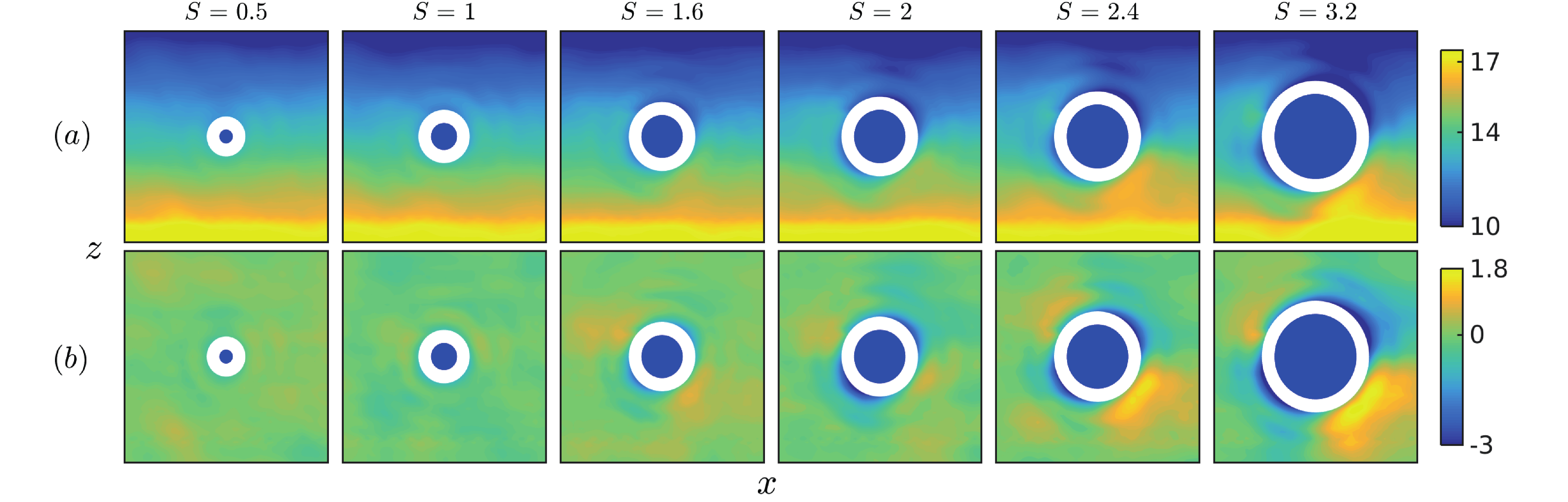}}
\caption{($a$)  Cross-sections $P(x,0,z)$ around the intruder, centered at the origin, for an intruder at $z_I=15$. The blue circle (diameter $d_i$) corresponds to the intruder. The edge of the white circle (diameter $d_i + d_b$) corresponds to the position of the first layer of bulk particles. ($b$) Cross-sections $P_L(x,0,z)$, where $P_L = P - P_H$, around the intruder at $z_I=15$.}
\label{fig:P_S_and_Pd_S}
\end{figure*}

A linear spring-dashpot model~\cite{CundallStrack1979, weinhart2013b} with linear elastic and linear dissipative contributions is used for the normal forces between particles. The restitution coefficient for collisions $r_c = 0.1$ and the contact duration $t_c = 0.005$. This results in a different stiffness depending on the particle size. We verified that our findings are not the result of this difference in stiffness nor the dependence on $r_c$ and $t_c$. The friction coefficient for contacts between bulk particles $\mu_{bb}$ and between bulk and intruder particles $\mu_{bi}$ equals 0.5, unless otherwise stated. 

We place three identical intruders in the flow at vertical positions $z_I=5$, $15$ and $23$ (see Fig.~\ref{fig:setup}). Each intruder is attached to a spring~\cite{guillard2016}, which applies a vertical force $F_{sp}= -k(z_i - z_I)$ proportional to the vertical distance between the intruder position $z_i$ and its corresponding $z_I$. Here $k=20$ is the spring stiffness. We also simulate $k=\infty$ by fixing the intruder at $z_i = z_I$. Our findings are independent of $k$, so unless stated otherwise all data reported are for $k=20$. We do not discuss the data for $z_I=5$ because the intruder experiences boundary effects, likely due to layering near the bed, as reported in~\cite{weinhart2013b}. 

The net contact force $F_{c}$ on an intruder can be determined in two ways: (i) Through the force balance $- F_{c} + F_{sp} - F_{g_z} = 0$, where $F_{sp}$ is computed from the intruder's average vertical position, and $F_{g_z} = \rho_p g_z V_i$ is the positively defined gravity force, with $V_i = \frac{4}{3}\pi (d_i/2)^3$ the intruder volume; (ii) By using the force balance $F_{c} = F_{n_z} + F_{t_z}$, with $F_{n_z}$ and $F_{t_z}$ the vertical normal and tangential contact forces, respectively. We verified that both methods give the same answer.


Applying coarse-graining (CG)~\cite{weinhart2013b, tunuguntla2016b, goldhirsch2010}, after a steady state has been reached, we obtain time-averaged 3D continuum fields for $\nu$ the local solids fraction, and $\sigma$ the stress tensor, which satisfy the conservation laws. The CG-width is chosen of the order of the particle diameter $w=d_b$ to achieve both rather smooth fields and independence of the fields on $w$~\cite{tunuguntla2016b}. We approximate the bulk solids fraction at the position of the intruder $\nu(x_i,y_i,z_i)=\nu_i=V_i/\tilde{V_{i}}$ using the ratio of the particle volume $V_i$ and the Voronoi volume $\tilde{V_{i}}$, which we obtain through 3D weighted Voronoi tessellation (math.lbl.gov/voro++; \cite{rycroft2009}). All error-bars (shaded areas) correspond to a 95\% confidence interval.

\emph{Results: Velocity Lag.---}Our first and most obvious finding is that intruders that have a size ratio larger than one ($S>1$) are positioned (on average) above $z_I$, thus with a non-zero and negative value of $F_{sp}$. Our second 
finding is that the downstream velocity $v_{xi}$ of an intruder with $S>1$, experiences a lag 
$\lambda_x=\langle v_{xi}(t) - v_x (z_i,t)\rangle$ 
with respect to the downstream velocity $v_x(z_i)$ of the bulk at height $z_i$. Figure~\ref{fig:SvsLag_springParticle}($a$) shows that a large intruder ($S>1$) lags ($\lambda_x < 0$), while a same sized intruder ($S=1$) experience no lag, within the fluctuations. Interestingly, but outside the scope of this study, for $S<1$, when the intruder is smaller than the bulk particles and sinks, $\lambda_x$ flips sign and becomes a velocity raise (increase). Figure~\ref{fig:SvsLag_springParticle}($b$) shows that the lag velocity increases at higher positions in the flow. 

Based on the derivation in the supplementary material we propose the following expression for the lag:
\begin{equation}\label{lambda1}
\lambda_x = \frac{1} {\pi d_b} \frac{1}{\eta}\frac{\Delta F(S)}{c(S)S}
\end{equation}
where $c(S)$ is a coefficient that potentially depends on $S$, $\eta$ is the granular viscosity, and $\Delta F$ is the unknown upslope-directed---in the negative $x$-direction---and size-ratio-dependent force responsible for the lag. The data in Fig.~\ref{fig:SvsLag_springParticle} provides us with the $S$ dependency of $\lambda_x$ and confirms the $1/\eta$ dependency predicted by Eq.~\eqref{lambda1}. Namely, we find a good fit of the data using 
\begin{equation}\label{lambda_fit}
\lambda_x =a(1/S - 1)/\eta 
\end{equation}
The dimensional fit parameter $a$ accounts for the $1/\pi d_b$ in Eq.~\eqref{lambda1}, as well as for $\Delta F$, which has dependencies that cannot be straightforwardly extracted from the data in our chute-flow geometry. If certain assumptions are made, which we can't verify in this geometry, the dimensional parameter $a$ can be made non-dimensional, as described in the supplementary material.

 Importantly, both the $S$-dependent data and the $z_i$-dependent data in Fig.~\ref{fig:SvsLag_springParticle} can be fitted with the same value for $a$. This fit also demonstrates that $\Delta F(S) / c(S) \propto 1- S$. Further support for the correct scaling of $\lambda_x$ is provided in Fig.~\ref{fig:SvsLag_springParticle}($c$), where a collapse of the data---except for outliers---is shown when plotting $\eta \lambda_x$ as a function of $S$, while Fig.~\ref{fig:SvsLag_springParticle}($d$) shows that all data fall on a line with slope 1.0 when plotting $\eta \lambda_x$ as a function of $a(1/S - 1)$. 

\emph{Pressure.---}We look for the origin of the lag in the pressure field around the intruder, where $P=\mathrm{Tr}(\sigma)/3$. Figure~\ref{fig:P_S_and_Pd_S}($a$) shows the cross-section $P(x, 0, z)$ for different size ratios. For $S\leq1$ the pressure is (almost) hydrostatic, i.e., $P\approx P_H = \nu \rho_p g_z (h-z)$, with $\nu \approx 0.577$. A hydrostatic pressure $P_H$, with very little variation in the solids fraction as a function of height, is characteristic for the bulk of this type of flow~\cite{weinhart2012b}. For $S>1$, $P$ deviates from $P_H$, and a strong anisotropy manifests itself with a high pressure region at the bottom-front side of the intruder. Pressure variations of lower magnitude also appear around the intruder. This demonstrates that the presence of a large particle modifies the local pressure around it. Although it is known that pulling an object through a granular medium affects the local pressure~\cite{guillard2014, guillard2015}, recall that here the intruder is not actively pulled but fixed by a spring in the $z$ direction, while it can freely flow in the $x$-$y$ plane.

In order to isolate the non-hydrostatic effects in the pressure we study $P_L = P - P_H$. Figure~\ref{fig:P_S_and_Pd_S}($b$) shows that for $S \leq 1$ $P_L$ is zero, within the fluctuations, while $P_L$ increases for $S>1$ and is characterized by positive regions (over-pressure) in the lower right and upper left quadrants, and negative regions in the lower left and upper right quadrants. It seems reasonable now to correlate the lift force and the velocity lag to this non-hydrostatic pressure.

\begin{figure*}
\resizebox{\textwidth}{!}{\includegraphics{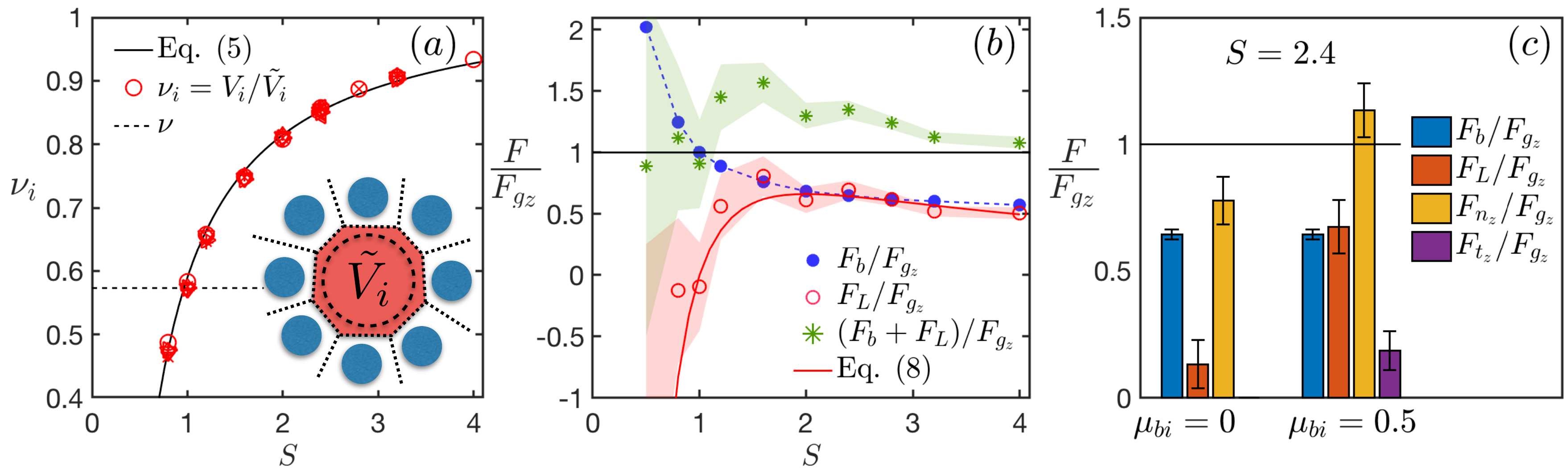} }
\caption{($a$) Local intruder solids fraction $\nu_i$ versus $S$. Different (almost collapsing) symbols correspond to intruders with $\mu_{bi}=0.5$, $\mu_{bi}=0$, $z_I=15$, $z_I=23$, $\theta = 22\degree, 23\degree, 24\degree, 25\degree$, and $26\degree$, $k=20$ and $k=\infty$. Solid line corresponds to Eq.~\eqref{nu_i} with $c=1.2$. The schematic depicts the Voronoi volume $\tilde{V_i}$ (dotted octagon) of the intruder (dashed circle). ($b$) The forces $F_b$, $F_L$ and $F_L + F_b$, normalized by $F_{g_z}$, for $z_I=23$. The dashed blue line is Eq.~\eqref{Fb} with $\nu_i$ from (a), while the solid red line is a fit of Eq. \eqref{eq:FL_saff} with $a=0.24$ and $b=130.0$. ($c$) The forces $F_b$, $F_L$, $F_{n_z}$ and $F_{t_z}$, normalized by $F_{g_z}$, for $S=2.4$, $\mu_{bi}=0$ and 0.5, at $z_I=15$.} 
\label{fig:awesome_plot}
\end{figure*}

\emph{Granular Buoyancy and Lift Force.---}{
Now that we have found indications that the velocity lag is linked to the local non-hydrostatic pressure field $P_L$, we proceed to calculate the lift force $F_L$ similar to the way we obtained $P_L$, i.e., by subtracting the granular buoyancy force $F_b$, that originates from $P_H$, from the net contact force on the intruder: $F_L = F_{c} - F_b$.} Different definitions for granular buoyancy forces exist~\cite[e.g][]{tripathi2011, guillard2016}, but here we derive a new definition that proofs to be essential to this study. Taking inspiration from~\cite{tripathi2011} and using our approximation $\nu(x_i,y_i,z_i)=\nu_i$ for the solids fraction at the intruder position, we integrate $P_H$ over the surface $\tilde{A_{i}}$ of $\tilde{V_{i}}$. With the divergence theorem we find:
\begin{equation}
\begin{split}
F_b =& \int_{\tilde{A_{i}}} P_H \textbf{n} \cdot \mathbf{e_z} \,d \tilde{A_{i}}      = \nu \rho_p g_z  \int_{ \tilde{V_{i}} } d  \tilde{V_{i}} 
       = \nu \rho_p g_z  \tilde{V_{i}} 
\end{split}
\end{equation}
Here $\textbf{n}$ is the normal outward vector to $\tilde{A_{i}}$ and $\mathbf{e_z}$ is the upward unit vector. 
Substituting $ \tilde{V_{i}}  = V_{i} / \nu_i $ we obtain:
\begin{equation}\label{Fb}
F_b = \frac{\nu}{\nu_i} \rho_p g_z V_i
\end{equation}
Effectively this is a generalized Archimedes principle at the particle level defined through an effective density that is equal to mass of the particle divided by its Voronoi volume. Figure~\ref{fig:awesome_plot}($a$) shows that the measured $\nu_i$ strongly depends on $S$ and is bigger than the bulk solids fraction $\nu$ for $S>1$. This means that a larger intruder occupies a larger fraction of its Voronoi volume. The data for $\nu_i$ can be fitted by:
\begin{equation}\label{nu_i}
\nu(x_i,y_i,z_i)=\nu_i=(\nu-1) S^c +1 ~,
\end{equation}
with $c=-1.2$ and $\nu = 0.577$.

The ratio $\nu / \nu_i$ in $F_b$ (Eq.~\eqref{Fb}) has a crucial consequence, namely that for $S>1$ the buoyancy force will be less than the gravity force $F_{g_z} = \rho_p g_z V_i$ acting on the particle. This can be seen in Fig.~\ref{fig:awesome_plot}($b$) where $F_b/F_{g_z} < 1$ for $S>1$. When $S=1$, $\nu$ equals $\nu_i$, and the buoyancy force balances $F_{g_z}$. In the limit of $S\rightarrow\infty$, we have that $\nu_i\rightarrow1$ and thus $F_b$ corresponds to the buoyancy force in a fluid with density $\rho=\nu \rho_p$. This generalized buoyancy force differs from the classical Archimedean buoyancy definition $F_b=\nu \rho_p g_z V_i$ in a granular fluid, which has two problems: it is independent of $S$, and more critically, predicts that $F_b < F_{g_z}$ if $S=1$. 

Using the new definition for $F_b$ we can determine the lift force $F_L = F_{c} - F_b$, with $F_{c} = F_{n_z} + F_{t_z}$. Figure~\ref{fig:awesome_plot}($b$) shows that $F_L/F_{g_z}$ is approximately zero for $S=1$, increases rapidly for $S>1$ and tends to a finite value above $S=2$. The plot of $(F_b + F_L )/ F_{g_z}$ in Fig.~\ref{fig:awesome_plot}($b$) shows that there is an optimal size ratio for segregation, in agreement with experimental findings~\cite{golick_daniels2009} and predictions~\cite{gray_ancey2011}.

\emph{Saffman Lift Force.---}Here we investigate the relation between the velocity lag of the intruder and the lift force it experiences. Such a relation is known to exist for suspended particles in a fluid: The Saffman lift force on a particle with diameter $d_i$ suspended in a fluid of density $\rho_f$ and viscosity $\eta_f$ is found to scale with the velocity lag with respect to the surrounding fluid~\cite{saffman1965, stone2000}:
\begin{equation}\label{eq:saffman}
F_{\text{Saffman}} = -1.615\, \sqrt{\eta_f | \dot{\gamma} |  \rho_f}  \lambda_x d_i^2 \, \mathrm{sgn}(\dot{\gamma}),
\end{equation}
where $\dot{\gamma} = \partial_z v_x(z_i)$ is the shear-rate. \citet{saffman1965} derived this relation taking the fluid properties in the absence of the particle and considered the limit:
\begin{equation}\label{saffmanlimits}
\frac{\rho_f\lambda_x d_i}{2\eta_f} \ll \left(\frac{\rho_f|\dot{\gamma}| d_i^2}{4\eta_f}\right)^{0.5}  \ll 1
\end{equation}
where the first term is the Reynolds number for the velocity lag $\mathcal{R}_{\lambda_x}$ and the second term is the  square root of the shear-rate Reynolds number $\mathcal{R}_{\dot{\gamma}}$. 
Note that for a granular fluid we can write $\mathcal{R}_{\dot{\gamma}}^{0.5} = I_\theta S / (2\sqrt{\mu})$, if we substitute the granular viscosity $\eta = \mu P |\dot{\gamma}|^{-1}$ and shear rate $|\dot{\gamma}|  =  I_\theta d_b^{-1}\sqrt{P/\rho_p}$, with $I_{\theta}$ the inertial number~\cite{jop2006constitutive}, and $\mu = \tan\theta$ the bulk friction.

Equation~\eqref{saffmanlimits} physically corresponds to a flow around an intruder that is locally governed by viscous effects ($\mathcal{R}_{\lambda_x} \ll 1$), but away from the intruder by inertial effects ($\mathcal{R}_{\lambda_x} \ll \mathcal{R}_{\dot{\gamma}}^{0.5}$). The derivation of the Saffman lift force is not valid when the inertia starts to dominate the local flow around the intruder,  and hence the validity is constrained to $\mathcal{R}_{\dot{\gamma}}^{0.5}  \ll 1$. Whether Eq.~\eqref{saffmanlimits} is valid for dense granular flows in general remains to be seen, nonetheless it is valid for our current system; we find $\mathcal{R}_{\lambda_x} =\mathcal{O}(10^{-4})$ using $\rho_f=\nu\rho_p$ and measuring $\eta$ from CG-fields in absence of the intruder, while $\mathcal{R}_{\dot{\gamma}}^{0.5}=I_{22\degree} S / (2\sqrt{\mu})= \mathcal{O}(10^{-1})$ using $I_{22\degree} = 0.050$. 



\emph{Granular Saffman Lift Force.---}In order to test if a Saffman-like relation exists between $F_L$ and $\lambda_x$ we define
\begin{equation}\label{eq:FL_saff}
F_L= -b\sqrt{\eta |\dot{\gamma}|\rho}\lambda_x d_i^2 \, \mathrm{sgn}(\dot{\gamma})
\end{equation}
analogous to Eq.~(\ref{eq:saffman}). 
Here $b$ a dimensionless coefficient that accounts for unknown dependencies, $\lambda_x=a(1/S-1)/\eta$ corresponding to Eq.~\eqref{lambda_fit}, and $\rho = \nu\rho_p$. Using $\eta^{-1}\sqrt{\eta |\dot{\gamma}|\rho}$ = $I_\theta (d_b \sqrt{\mu})^{-1}$, Eq.~\eqref{eq:FL_saff} can be written as:
\begin{equation}\label{eq:FL_saff2}
F_L= -ab I_\theta\mu^{-0.5}  (1/S - 1)d_i^2 d_b^{-1}   \, \mathrm{sgn}(\dot{\gamma}),
\end{equation}
demonstrating that the lift force is independent of the flow depth, since $I_\theta$ and $\mu$ are constant in a chute flow. We verify that $F_L$ is indeed independent of depth (see the supplementary material), in agreement with the findings of~\citet{guillard2016}. 

We fit Eq.~\eqref{eq:FL_saff} to the data of $F_L$ in Fig.~\ref{fig:awesome_plot}($b$), using the value for $a$ obtained from the fit in Fig.~\ref{fig:SvsLag_springParticle}, and find that it captures the data well. Subsequently, using the same value for $a$, and the value for $b$ obtained from the fit to $F_L$ in Fig.~\ref{fig:awesome_plot}($b$), we fit Eq.~\eqref{eq:FL_saff} to the lift force measured as a function of depth in the supplementary material.  This demonstrates that Eq.~\eqref{eq:FL_saff} is the correct scaling between the lift force, size ratio, viscosity and velocity lag at constant inclination angle in a chute flow. The fact that this scaling is Saffman-like suggests that inertial effects could lie at the origin of the segregation of large particles in dense granular flows with pressure- and velocity gradients in the limit of low large-particle concentrations.

To provide further support for our finding that the generalized buoyancy force does not support the weight of a large intruder ($S>1$) we set the intruder-bulk friction $\mu_{bi}$ to zero and find that $F_L$ is reduced, as shown in Fig.~\ref{fig:awesome_plot}($c$). Critically, this leads to a large none-frictional intruder \emph{sinking} instead of rising, as found recently also experimentally: lower-friction particles sink below higher-friction particles in mono-disperse granular flows~\cite{gillemot2017}. Since the net contact force $F_{c}= F_{n_z} + F_{t_z}$ on the intruder is lower than $F_{g_z}$, the buoyancy $F_b$ must also be less than $F_{g_z}$. Note that in Fig.~\ref{fig:awesome_plot}($c$) the spring force brings the force balance back to zero: $F_{sp} - F_{g_z} + F_{c} = F_{sp} - F_{g_z} + F_b + F_L = 0$. Interestingly, the lift force does not completely disappear, indicating it should have both a geometric and frictional component. We verified that $P_L$ is reduced but does not disappear for frictionless particles. 




\emph{Conclusions.---}We report that a single large particle in a dense granular flow is surrounded by an anisotropic, non-hydrostatic pressure field. This coincides with our observations of a velocity lag $\lambda_x$ and a lift force $F_L$, coupled through a Saffman-like relation, Eq.~\eqref{eq:FL_saff}, causing the particle to rise against gravity. These findings suggest that the mechanism of squeeze expulsion~\cite{savage_lun1988}---which has been used to qualitatively explain the segregation of large particles in dense granular flows---is the granular equivalent of the Saffman effect; an inertial lift force in an otherwise strongly viscous bulk flow~\cite{saffman1965, stone2000}. 

A possible physical interpretation of the Saffman effect for a granular fluid could be that in our mostly viscous and slow flow, but with a finite, considerable inertial number, a large intruder disturbs the local (Bagnold) flow profile. Because the bulk inertial effects, which are proportional to the strain-rate, are not negligible, the rheology driven by the velocity gradient---associated with the inertially generated, but perturbed velocity field---produces an anisotropy of the pressure field, which creates both the lift force and the drag force responsible for the velocity lag.

The decomposition of the contact force on the intruder into a lift force and generalized buoyancy force is essential to the preceding analysis. Moreover, it provides a physical explanation for the sinking of very large intruders~\cite{thomas2000, felix2004}, as well as for the optimal size ratio for segregation~\cite{golick_daniels2009, gray_ancey2011} and the unexplained trend of $F_{c}(S)$ in Fig.~6 of~\cite{guillard2016}. Namely, if we consider the limit of Eq.~\eqref{eq:FL_saff} at large size ratios, we see that the lag approaches a constant value, while the buoyancy force approaches a fluid buoyancy with density $\rho = \nu \rho_p$. Gravity will then outgrow the total upward force and the particle will sink. 

Further studies could address the following questions: If inertial effects indeed lie at the origin of size segregation of large intruders at low large-particle concentration, they could potentially also play a role in slow, dense, polydisperse granular flows with more than one intruder. Thus, the variation of the lift force when the large-particle concentration increases could be investigated. Furthermore, in order to validate the Saffman relation for granular flows changing the stress gradient in the flow would be necessary. This can be done by using other geometries, for example, such as the one used by~\citet{guillard2016}. Last but not least, the reported sinking of a large intruder with zero intruder-bulk friction $\mu_{bi}$ hints at the importance of particle properties.

Drag forces on a free-flowing object in granular media, in contrast to a dragged object, have received little attention~\cite{tripathi2011}. Our findings suggest that the Stokesian drag, found by \citet{tripathi2011} for a heavy sinking mono-disperse intruder, plays an important role in the rising of large intruders (see the supplementary material). A continued effort to determine all drag forces acting on free-flowing particles is important for the rheology of granular flows in general, but foremost because drag is a cornerstone of models for particle-size segregation in dense granular flows.

In order to unify Eq.~\eqref{eq:FL_saff} with the scaling laws found by \citet{guillard2016} and develop a multi-scale model for the segregation of large intruders in dense granular flows the lag will have to be expressed in terms of $\lambda_x = f(\partial P/\partial z, \partial |\tau| / \partial z, \dot{\gamma}, \partial \dot{\gamma}/\partial z)$, where $\tau$ is the shear stress. This is far from trivial and ongoing work: The dependency of all variables on $z$ and $\theta$ is very weak and the range of accessible pressure gradients, inertial numbers, etc., is very limited in steady state chute flows (inclination angles that are too large lead to accelerating flows, whereas too small angles lead to stopping of the flow \cite{weinhart2012b, weinhart2013}). To demonstrate the dependencies more convincingly, one should disentangle pressure and tangential stress and show that the Saffman-like relation still holds. In order to do so, a completely different flow geometry needs to be considered, which, however, goes beyond the scope of the present study. Finally, for a formal proof that a Saffman-like relation holds in granular fluids, the analytical derivation by Saffman could be repeated for a granular rheology.



\begin{acknowledgments}
\emph{Acknowledgments.---}KV and MS contributed equally to this study. The authors acknowledge Chris G. Johnson for suggesting the Saffman effect, Fran\c{c}ois Guillard for helpful email correspondence and commenting on the manuscript, and the referees for their critical help improving the manuscript. KV is also grateful to S.C. van Keulen for many fruitful discussions. The authors acknowledge support from the Swiss National Science Foundation grant No. 200021\_149441 $/$ 1 and the Dutch Technology Foundation STW grant STW-Vidi Project 13472. \end{acknowledgments}

%

\end{document}


\widetext


\title{Segregation of large particles in dense granular flows: A granular Saffman effect? --- Supplementary material}


\author{K. van der Vaart}
\affiliation{Environmental Hydraulics Laboratory, \'{E}cole Polytechnique F\'{e}d\'{e}rale de Lausanne, \'{E}cublens, 1015 Lausanne, Switzerland}
\affiliation{Multi-Scale Mechanics, ET and MESA+, University of Twente, P.O. Box 217, 7500AE Enschede, The Netherlands}

\author{M. P. van Schrojenstein Lantman}
\affiliation{Multi-Scale Mechanics, ET and MESA+, University of Twente, P.O. Box 217, 7500AE Enschede, The Netherlands}

\author{T. Weinhart}
\affiliation{Multi-Scale Mechanics, ET and MESA+, University of Twente, P.O. Box 217, 7500AE Enschede, The Netherlands}

\author{S. Luding}
\affiliation{Multi-Scale Mechanics, ET and MESA+, University of Twente, P.O. Box 217, 7500AE Enschede, The Netherlands}

\author{C. Ancey}
\affiliation{Environmental Hydraulics Laboratory, \'{E}cole Polytechnique F\'{e}d\'{e}rale de Lausanne, \'{E}cublens, 1015 Lausanne, Switzerland}

\author{A.R. Thornton}
\affiliation{Multi-Scale Mechanics, ET and MESA+, University of Twente, P.O. Box 217, 7500AE Enschede, The Netherlands}

\date{\today}

\newcounter{defcounter}
\setcounter{defcounter}{0}
\newenvironment{Sequation}{
\addtocounter{equation}{-1}
\refstepcounter{defcounter}
\renewcommand\theequation{S\thedefcounter}
\begin{equation}}
{\end{equation}}

\maketitle

\section{Inclination angle dependence of the lift force}

\begin{figure}
\resizebox{0.85\columnwidth}{!}{  \includegraphics{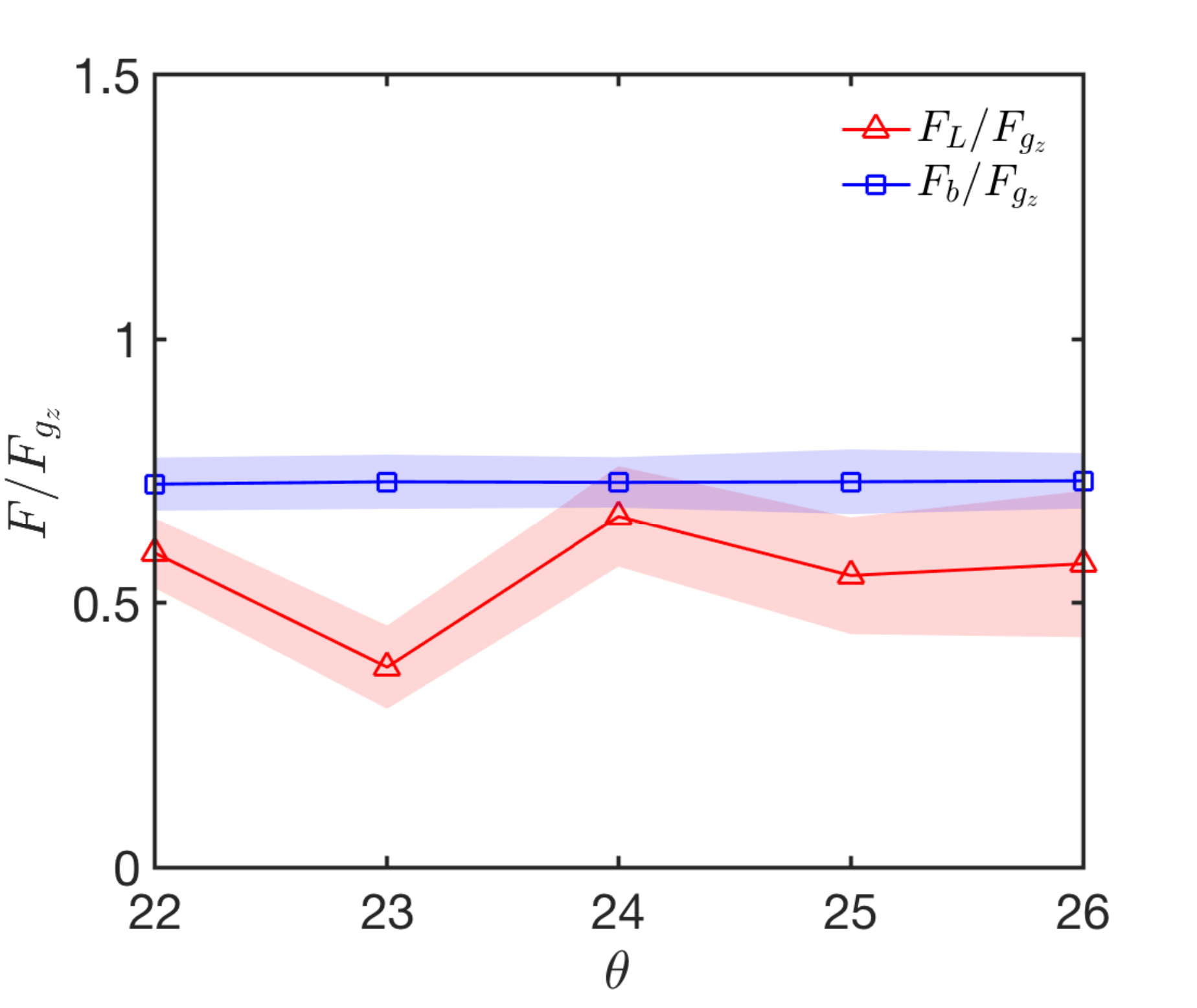}}
\caption{The lift force $F_L$ (red triangles) and the buoyancy force $F_b$ (blue squares) as a function of the chute inclination angle $\theta$, for $S=2.4$. The data are normalized by the vertical component of the gravity force on the intruder $F_{g_z}$. }
\label{fig:f_theta}
\end{figure}

In this supplementary material we show the dependence of the lift force $F_L$, as well as the buoyancy $F_b$, on the inclination angle $\theta$ of the chute, for $S=2.4$. These data are plotted in Fig.~\ref{fig:f_theta} where we see that the lift force and the buoyancy force are independent of the inclination angle, within the fluctuations.

\section{Horizontal force balance and velocity lag}
In this supplementary material we introduce a scaling for the lag velocity $\lambda_x$ based on the horizontal force balance. Note that by our definition the lag velocity is negative, i.e., the intruder is moving slower than the bulk material at its height. The aim is to show what parameter dependencies are present in the fitting parameter $a$ in Eq.~(2), which is 
\begin{Sequation}
\label{lambda_fit}
\lambda_x =a(\frac{1}{S} - 1)/\eta,
\end{Sequation}where $\eta = \mu P / \dot{\gamma}$ is he granular viscosity, with $\mu$ the bulk friction, $P$ the pressure, and $\dot{\gamma} = \partial_z v_x$ the shear rate.

When the size ratio equals one $(S=1)$ we have the following horizontal force balance on the intruder:
\begin{Sequation}
F_r(S) + F_{g_x}(S) = 0,
\end{Sequation}where $F_{g_x} = \rho_p g_x V_i$ is the horizontal component of the gravitational force (with $g_x = \sin\theta$), and $F_r$ is the (negative) net horizontal contact force, resembling a "fricional" buoyancy force caused by the shear stress, that cancels gravity.

When the size ratio becomes larger than one $(S>1)$ and the intruder starts to experience a velocity lag $\lambda_x$, we propose that the horizontal force balance can be written as:
\begin{Sequation}\label{balance2}
F_d(S, \lambda_x) + F_r(S) + F_{g_x}(S) = 0
\end{Sequation}where the "frictional" buoyancy force $F_r$ become bigger than the downslope gravity force $F_{g_x}$, which causes a lag that is damped by $F_d$, a Stokesian-like drag working in the same direction as $F_{g_x}$ with a dependence on the lag velocity. Note that combined, $F_d$ and $F_r$ form the contact force, in the $x$-direction, experienced by the intruder,
\begin{Sequation}
F_{c_x} = F_d + F_r.
\end{Sequation}In steady state, when the lag is constant in time, there is likely a drag force proportional to the lag velocity, acting in the opposite direction to it, to prevent the intruder from accelerating. The granular Stokes drag introduced by~\citet{tripathi2011} is a good candidate for this role:
\begin{Sequation}\label{F_d}
F_d = - c(S) \pi \eta \lambda_x d_i.
\end{Sequation}Here $c(S)$ is a coefficient, and $d_i$ is the diameter of the intruder. \citet{tripathi2011} obtained this drag force for a heavy (higher density) mono-disperse intruder in a chute flow, where they measured the vertical velocity of the sinking intruder. Note that a dependence of $F_d$ on $S$ is a possibility, because the drag force appears to be a function of the volume fraction~\cite{tripathi2011} and locally the experienced volume fraction by the intruder changes as function of $S$ (see Fig~4($a$)).

The sum of $F_r$ and $F_{g_x}$ in Eq.~\eqref{balance2}, which we denote as $\Delta F(S) =   F_r(S) + F_{g_x}(S)$ (analogous to $F_L$ in the main-text), can be understood as the upslope directed---in the negative $x$-direction---force causing the intruder to lag. Swapping $F_d$ for $-\Delta F(S)$ in Eq.~\eqref{F_d} and using $d_i = S d_b$ we obtain an expression for the lag:
\begin{Sequation}\label{lambda1}
\lambda_x = \frac{1}{ \pi d_b}\frac{1}{\eta} \frac{\Delta F(S)}{c(S) S},
\end{Sequation}where the first factor is constant, the inverse viscosity represents the second factor, and the $S$-dependence is condensed into the third factor. Hence, the dimensional fitting parameter $a$ in Eq.~(2) accounts for the dependency on the bulk particle diameter and unknown dependencies of $c(S)$ and $\Delta F(S)$.

In order to non-dimensionalize Eq.~\eqref{lambda1} an assumption has to be made for $\Delta F(S)$. If we would assume that $\Delta F(S)$ is proportional to the shear gradient $\frac{\partial \tau}{\partial z}$ and the volume of the particle $V_i$ we can write:
\begin{Sequation}
\label{assumption}
\Delta F(S) = f(S)\frac{\partial \tau}{\partial z}V_i,
\end{Sequation}where $f(S)$ is some $S$-dependent function. This would yield for the lag velocity: 
\begin{Sequation}
\lambda_x = \frac{V_i}{\pi d_b}\frac{\partial \tau}{\partial z} \frac{1}{\eta} \frac{f(S)}{c(S) S} = v_b n(S),
\end{Sequation}where $v_b = \frac{\partial \tau}{\partial z} \frac{d_i^2}{\eta} $ and $n(S) = \frac{f(S)}{6c(S)}$. This would render the $S$-dependent term $n(S)$ dimensionless, while the factor $v_b$ is a situation dependent constant with units of velocity, proportional to the viscosity, the gradient in shear stress and the particle diameter. The connection between $v_b$ and the constant coefficient $a$ in Eq.~\eqref{lambda_fit} can be found by writing $v_b = aS^2$. This reveals that $a \propto \frac{\partial \tau}{\partial z} d_b^2$. However, Eq.~\eqref{assumption} is an assumption that we are not willing to make, so that we use instead Eq.~\eqref{lambda1} in the main text.

\section{Depth dependence of the lift force}
\begin{figure}[b!]
\vspace{3mm}
\resizebox{0.9\columnwidth}{!}{ \includegraphics{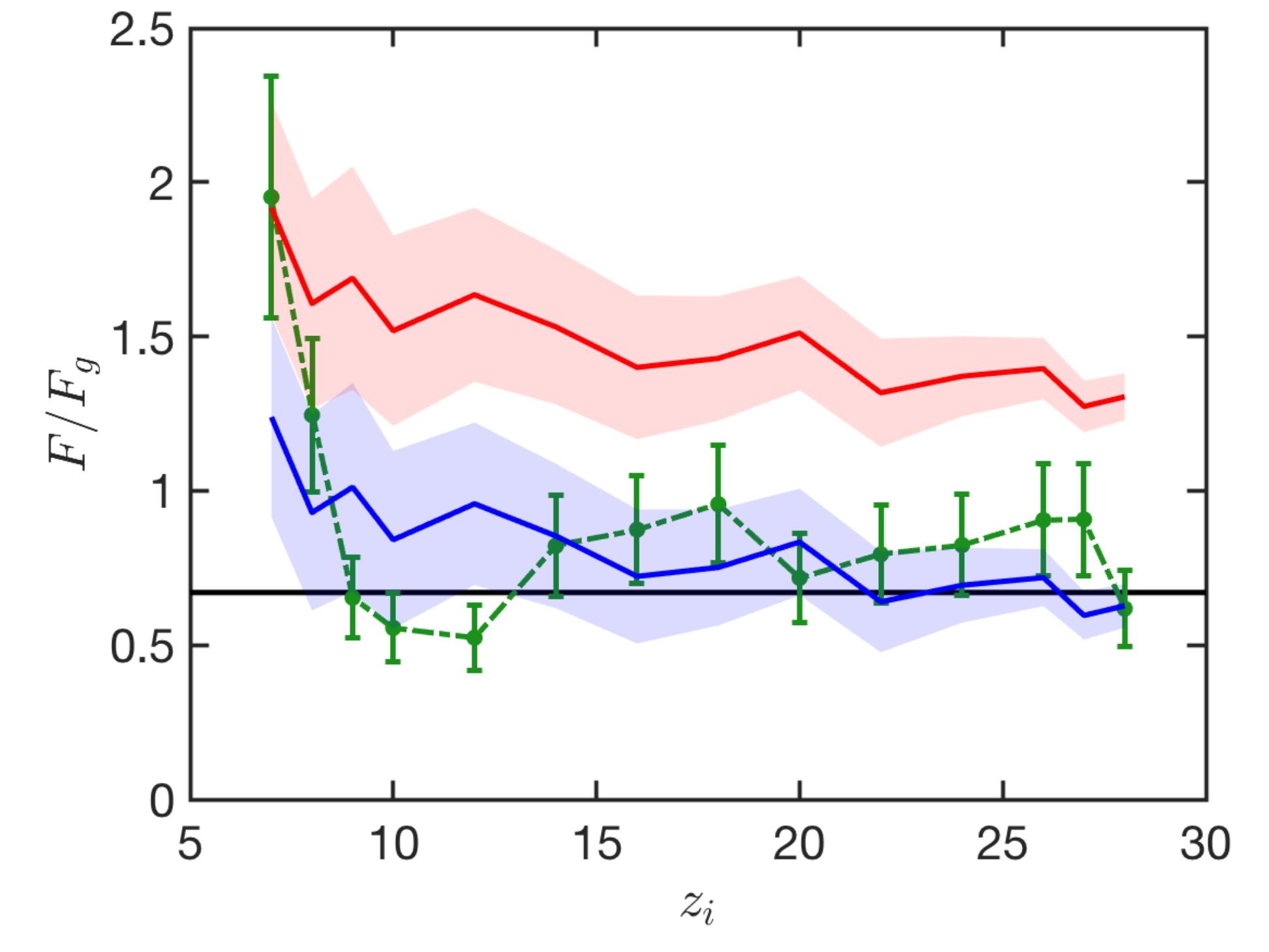}}
\caption{The net contact force $F_c$ (red line) and the lift force $F_L$ (blue line) on an intruder, as a function of depth, for S=2.4, normalized by the gravity force. The solid black line is a fit of Eq.~(7) using the functional form for the lag $\lambda_x = a(1/S - 1)/\eta$, with $a=0.24$ and $b=130.0$, while the dashed line uses the raw velocity lag data from Fig~2($a$). Near the bed a boundary effect occurs, likely due to layering~\cite{weinhart2013b}.}
\label{fig:f_z}
\end{figure}

In this supplementary material we show the dependence of the lift force $F_L$, as well as the total upward force $F_{tot} = F_L + F_b$ on the depth of the intruder. These data are plotted in Fig.~\ref{fig:f_z} where we see that the lift force and total force are independent of the depth. This is in agreement with the findings reported by \citet{guillard2016}. There is an increase of both forces close to the bed, but we attribute this to a boundary effect where the intruder particle experiences a greater force due to layering of particles near the bed, as reported by~\citet{weinhart2013b}.

Two fits of the lift force model, Eq.~(7), are shown in Fig.~\ref{fig:f_z}, one using the functional form of the lag $\lambda_x = a(1/S - 1)/\eta$ and a second using the raw data for the velocity lag from Fig.~2($b$).

%